\documentclass[apjl]{emulateapj}
\usepackage{comment}
\usepackage{ifthen}

\newcommand{\forloop}[5][1]%
{%
\setcounter{#2}{#3}%
\ifthenelse{#4}%
	{%
	#5%
	\addtocounter{#2}{#1}%
	\forloop[#1]{#2}{\value{#2}}{#4}{#5}%
	}%
	{%
	}%
}%

\newcommand{\ctbd}[1]{}

\newcommand{\lc}{light curve}
\newcommand{\lcs}{light curves}
\newcommand{\Lc}{Light curve}

\newcommand{\band}[1]{\ensuremath{#1}~band}

\newcommand{\kms}{\ensuremath{\rm km\,s^{-1}}}
\newcommand{\ms}{\ensuremath{\rm m\,s^{-1}}}

\newcommand{\gcmc}{\ensuremath{\rm g\,cm^{-3}}}
\newcommand{\ergscmsq}{\ensuremath{\rm erg\,s^{-1}\,cm^{-2}}}

\newcommand{\vsini}{\ensuremath{v \sin{i}}}
\newcommand{\feh}{\ensuremath{\rm [Fe/H]}}

\newcommand{\vmac}{\ensuremath{v_{\rm mac}}}
\newcommand{\vmic}{\ensuremath{v_{\rm mic}}}

\newcommand{\rsun}{\ensuremath{R_\sun}}
\newcommand{\msun}{\ensuremath{M_\sun}}
\newcommand{\lsun}{\ensuremath{L_\sun}}

\newcommand{\rstar}{\ensuremath{R_\star}}
\newcommand{\mstar}{\ensuremath{M_\star}}
\newcommand{\lstar}{\ensuremath{L_\star}}

\newcommand{\teffstar}{\ensuremath{T_{\rm eff\star}}}
\newcommand{\rhostar}{\ensuremath{\rho_\star}}
\newcommand{\loggstar}{\ensuremath{\log{g_{\star}}}}

\newcommand{\rpl}{\ensuremath{R_{p}}}
\newcommand{\mpl}{\ensuremath{M_{p}}}

\newcommand{\rhopl}{\ensuremath{\rho_{p}}}

\newcommand{\arstar}{\ensuremath{a/\rstar}}
\newcommand{\zrstar}{\ensuremath{\zeta/\rstar}}

\newcommand{\rjup}{\ensuremath{R_{\rm J}}}
\newcommand{\mjup}{\ensuremath{M_{\rm J}}}

\newcommand{\reffigl}[1]{Figure~\ref{fig:#1}}
\newcommand{\refsecl}[1]{\mbox{Section \ref{sec:#1}}}

\newcommand{\reftabl}[1]{Table~\ref{tab:#1}}

\newcommand{\flwof}{\mbox{FLWO 1.2\,m}}

\newcommand{\hatcurfield}{364}                                         
\newcommand{\hatcurCCra}{\ensuremath{08^{\mathrm h}15^{\mathrm m}47.99{\mathrm s}}}                                  
\newcommand{\hatcurCCdec}{\ensuremath{+05{\arcdeg}50{\arcmin}12.2{\arcsec}}}                                 
\newcommand{\hatcurCCgsc}{GSC~0208-00722}                              
\newcommand{\hatcurCCtassmv}{10.419}                                   
\newcommand{\hatcurCCtwomassJmag}{\ensuremath{9.442\pm0.026}}          
\newcommand{\hatcurCCtwomassHmag}{\ensuremath{9.220\pm0.028}}          
\newcommand{\hatcurCCtwomassKmag}{\ensuremath{9.151\pm0.023}}          
\newcommand{\hatcurLCdip}{\ensuremath{8.8}}                            
\newcommand{\hatcurLCrprstar}{\ensuremath{0.1134\pm0.0020}}            
\newcommand{\hatcurLCbsq}{\ensuremath{0.729_{-0.017}^{+0.014}}}        
\newcommand{\hatcurLCimp}{\ensuremath{0.854_{-0.010}^{+0.008}}}        
\newcommand{\hatcurLCzeta}{\ensuremath{30.96\pm0.45}}                  
\newcommand{\hatcurLCdur}{\ensuremath{0.0887\pm0.0015}}                
\newcommand{\hatcurLCingdur}{\ensuremath{0.0295\pm0.0025}}             
\newcommand{\hatcurLCP}{\ensuremath{2.810595\pm0.000005}}              
\newcommand{\hatcurLCPprec}{\ensuremath{2.8105951}}                    
\newcommand{\hatcurLCPshort}{\ensuremath{2.8106}}                      
\newcommand{\hatcurLCT}{\ensuremath{2455456.46561\pm0.00037}}          
\newcommand{\hatcurSMEiteff}{\ensuremath{6169\pm100}}                  
\newcommand{\hatcurSMEizfeh}{\ensuremath{+0.06\pm0.1}}                  
\newcommand{\hatcurSMEizfehshort}{\ensuremath{+0.06}}                   
\newcommand{\hatcurSMEilogg}{\ensuremath{4.18\pm0.06}}                 
\newcommand{\hatcurSMEivsin}{\ensuremath{2.9\pm0.5}}                   
\newcommand{\hatcurSMEivmac}{\ensuremath{4.63}}                        
\newcommand{\hatcurSMEivmic}{\ensuremath{0.85}}                        
\newcommand{\hatcurSMEiiteff}{\ensuremath{6304\pm88}}                  
\newcommand{\hatcurSMEiizfeh}{\ensuremath{+0.13\pm0.08}}                
\newcommand{\hatcurSMEiizfehshort}{\ensuremath{+0.13}}                  
\newcommand{\hatcurSMEiilogg}{\ensuremath{4.37\pm0.06}}                
\newcommand{\hatcurSMEiivsin}{\ensuremath{2.2\pm0.5}}                  
\newcommand{\hatcurSMEiivmac}{\ensuremath{4.84}}                       
\newcommand{\hatcurSMEiivmic}{\ensuremath{0.85}}                       
\newcommand{\hatcurTRESgamma}{\ensuremath{45.51\pm0.18}}               
\newcommand{\hatcurLBiz}{\ensuremath{0.1448}}                          
\newcommand{\hatcurLBiiz}{\ensuremath{0.3599}}                         
\newcommand{\hatcurLBii}{\ensuremath{0.1975}}                          
\newcommand{\hatcurLBiii}{\ensuremath{0.3689}}                         
\newcommand{\hatcurISOm}{\ensuremath{1.24\pm0.04}}                     
\newcommand{\hatcurISOmlong}{\ensuremath{1.242\pm0.041}}               
\newcommand{\hatcurISOr}{\ensuremath{1.21\pm0.05}}                     
\newcommand{\hatcurISOrlong}{\ensuremath{1.215\pm0.051}}               
\newcommand{\hatcurISOlogg}{\ensuremath{4.36\pm0.03}}                  
\newcommand{\hatcurISOlum}{\ensuremath{2.05\pm0.24}}                   
\newcommand{\hatcurISOmv}{\ensuremath{3.98\pm0.14}}                    
\newcommand{\hatcurISOage}{\ensuremath{1.0_{-0.5}^{+0.8}}}             
\newcommand{\hatcurISOMK}{\ensuremath{2.77\pm0.10}}                    
\newcommand{\hatcurRVK}{\ensuremath{88.1\pm3.3}}                       
\newcommand{\hatcurRVk}{\ensuremath{-0.006\pm0.015}}                   
\newcommand{\hatcurRVh}{\ensuremath{-0.027\pm0.034}}                   
\newcommand{\hatcurRVjitterA}{\ensuremath{6.3}}                        
\newcommand{\hatcurRVfitrmsA}{\ensuremath{6.7}}                        
\newcommand{\hatcurRVjitterB}{\ensuremath{2.1}}                        
\newcommand{\hatcurRVfitrmsB}{\ensuremath{12.3}}                       
\newcommand{\hatcurRVeccen}{\ensuremath{0.035\pm0.024}}                
\newcommand{\hatcurRVomega}{\ensuremath{252\pm84}}                     
\newcommand{\hatcurPPi}{\ensuremath{83.6\pm0.4}}                       
\newcommand{\hatcurPPlogg}{\ensuremath{2.99\pm0.04}}                   
\newcommand{\hatcurPPar}{\ensuremath{7.42\pm0.26}}                     
\newcommand{\hatcurPParel}{\ensuremath{0.0419\pm0.0005}}               
\newcommand{\hatcurPPrho}{\ensuremath{0.37\pm0.05}}                    
\newcommand{\hatcurPPmlong}{\ensuremath{0.711\pm0.028}}                
\newcommand{\hatcurPPrlong}{\ensuremath{1.340\pm0.065}}                
\newcommand{\hatcurPPmrcorr}{\ensuremath{0.07}}                        
\newcommand{\hatcurPPteff}{\ensuremath{1630\pm42}}                     
\newcommand{\hatcurPPtheta}{\ensuremath{0.035\pm0.002}}                
\newcommand{\hatcurPPfluxperi}{\ensuremath{1.72\pm0.16}}               
\newcommand{\hatcurPPfluxperidim}{\ensuremath{9}}                      
\newcommand{\hatcurPPfluxap}{\ensuremath{1.49\pm0.19}}                 
\newcommand{\hatcurPPfluxapdim}{\ensuremath{9}}                        
\newcommand{\hatcurPPfluxavg}{\ensuremath{1.59\pm0.16}}                
\newcommand{\hatcurPPfluxavgdim}{\ensuremath{9}}                       
\newcommand{\hatcurXsecondary}{\ensuremath{2455457.861\pm0.027}}       
\newcommand{\hatcurXsecdur}{\ensuremath{0.0895\pm0.0020}}              
\newcommand{\hatcurXsecingdur}{\ensuremath{0.0236\pm0.0077}}           
\newcommand{\hatcurXdist}{\ensuremath{193\pm8}}                        

\newcommand{\hatcur}{HAT-P-30}
\newcommand{\hatcurb}{HAT-P-30b}

\newcommand{\hatcurCCtassvi}{\ensuremath{0.527\pm0.12}}                  
\newcommand{\hatcurSMEversion}{ii}                                       
\newcommand{\hatcurSMEteff}{\ifthenelse{\equal{\hatcurSMEversion}{i}}{\hatcurSMEiteff}{\hatcurSMEiiteff}}
\newcommand{\hatcurSMEzfeh}{\ifthenelse{\equal{\hatcurSMEversion}{i}}{\hatcurSMEizfeh}{\hatcurSMEiizfeh}}
\newcommand{\hatcurSMEzfehshort}{\ifthenelse{\equal{\hatcurSMEversion}{i}}{\hatcurSMEizfehshort}{\hatcurSMEiizfehshort}}
\newcommand{\hatcurSMElogg}{\ifthenelse{\equal{\hatcurSMEversion}{i}}{\hatcurSMEilogg}{\hatcurSMEiilogg}}
\newcommand{\hatcurSMEvsin}{\ifthenelse{\equal{\hatcurSMEversion}{i}}{\hatcurSMEivsin}{\hatcurSMEiivsin}}
\newcommand{\hatcurSMEvmac}{\ifthenelse{\equal{\hatcurSMEversion}{i}}{\hatcurSMEivmac}{\hatcurSMEiivmac}}
\newcommand{\hatcurSMEvmic}{\ifthenelse{\equal{\hatcurSMEversion}{i}}{\hatcurSMEivmic}{\hatcurSMEiivmic}}

\newcommand{\hatcurisoshort}{YY}

\newcommand{\hatcurisocite}{yi:2001}
\newcommand{\hatcurlumind}{\arstar}
\newcommand{\hatcurjhkfilset}{ESO}
\def\lam{73.5^\circ}
\def\ulam{9.0^\circ}

\newboolean{emulateapj}
\setboolean{emulateapj}{true}

\newboolean{rvtablelong}
\setboolean{rvtablelong}{false}

\newboolean{astroph}
\setboolean{astroph}{true}

\shortauthors{Johnson et al.}
\shorttitle{\hatcur\lowercase{b}}
\ifthenelse{\boolean{emulateapj}}{
    \newcommand{\titledag}{$\dagger$}
}{
    \newcommand{\titledag}{\dagger}
}

\begin{document}

\title{\hatcur\lowercase{b}: A transiting hot Jupiter on a 
	highly oblique orbit \altaffilmark{\titledag}}

\author{
	John~Asher~Johnson\altaffilmark{1},
        J.~N.~Winn\altaffilmark{2},
	G.~\'A.~Bakos\altaffilmark{3},
	J.~D.~Hartman\altaffilmark{3},
	T.~D.~Morton\altaffilmark{1},
	G.~Torres\altaffilmark{3},
	G\'eza~Kov\'acs\altaffilmark{4},
	D.~W.~Latham\altaffilmark{3},
	R.~W.~Noyes\altaffilmark{3},
	B.~Sato\altaffilmark{5},
	G.~A.~Esquerdo\altaffilmark{3},
	D.~A.~Fischer\altaffilmark{6},
	G.~W.~Marcy\altaffilmark{7},
	A.~W.~Howard\altaffilmark{7},
        L.~A.~Buchhave\altaffilmark{8},
        G.~F\H{u}r\'{e}sz\altaffilmark{3},
	S.~N.~Quinn\altaffilmark{3},
        B.~B\'eky\altaffilmark{3},
	D.~D.~Sasselov\altaffilmark{3},
	R.~P.~Stefanik\altaffilmark{3},
	J.~L\'az\'ar\altaffilmark{9},
	I.~Papp\altaffilmark{9},
	P.~S\'ari\altaffilmark{9},
}
\altaffiltext{1}{California Institute of Technology,
  Department of Astrophysics, MC 249-17, Pasadena, CA 91125; NASA
  Exoplanet Science Institute (NExScI)} 

\altaffiltext{2}{Department of Physics, and Kavli Institute for
  Astrophysics and Space Research, Massachusetts Institute  
  of Technology, Cambridge, MA 02139}

\altaffiltext{3}{Harvard-Smithsonian Center for Astrophysics,
	Cambridge, MA; email: gbakos@cfa.harvard.edu}

\altaffiltext{4}{Konkoly Observatory, Budapest, Hungary.}

\altaffiltext{5}{Department of Earth and Planetary Sciences, 
	Tokyo Institute of Technology, 2-12-1 Ookayama, Meguro-ku, 
	Tokyo 152-8551}

\altaffiltext{6}{Department of Astronomy, Yale University, New Haven, CT}

\altaffiltext{7}{Department of Astronomy, University of California,
	Berkeley, CA}

\altaffiltext{8}{Niels Bohr Institute, Copenhagen University, DK-2100
Copenhagen, Denmark}

\altaffiltext{9}{Hungarian Astronomical Association, Budapest, 
	Hungary}

\altaffiltext{$\dagger$}{
	Based in part on observations obtained at the W.~M.~Keck
        Observatory, which is operated by the University of California
        and the California Institute of Technology. Keck time has been
        granted by NASA (N167Hr). Based in part on data collected at
        Subaru Telescope, which is operated by the National
        Astronomical Observatory of Japan.  
}

\begin{abstract}
\setcounter{footnote}{10}

We report the discovery of \hatcurb{}, a transiting exoplanet orbiting
the V=\hatcurCCtassmv\ dwarf star \hatcurCCgsc.  The planet has a
period $P=\hatcurLCP$\,d, transit epoch $T_c = \hatcurLCT$ (BJD), and
transit duration \hatcurLCdur\,d.  The host star has a mass of
\hatcurISOm\,\msun, radius of \hatcurISOr\,\rsun, effective temperature
\hatcurSMEteff\,K, and metallicity $\feh = \hatcurSMEzfeh$.  The
planetary companion has a mass of \hatcurPPmlong\,\mjup, and radius of
\hatcurPPrlong\,\rjup\ yielding a mean density of \hatcurPPrho\,\gcmc. 
We also present radial velocity measurements that were obtained
throughout a transit that exhibit the Rossiter-McLaughlin effect.  By
modeling this effect we measure an angle of $\lambda = \lam \pm \ulam$
between the sky projections of the planet's orbit normal and the star's
spin axis.  \hatcurb{} represents another example of a close-in planet
on a highly tilted orbit, and conforms to the previously noted pattern
that tilted orbits are more common around stars with $\teffstar \gtrsim
6250$~K.
\setcounter{footnote}{0}
\end{abstract}


\section{Introduction}
\label{sec:introduction}

The majority of known exoplanets have been discovered either by
detecting the gravitational pull of a planet on its central star using
Doppler spectroscopy, or by observing the small decrement in flux from
the star during a transit of its planet.  The two techniques are
complementary: several Doppler-detected planets have later been found
to transit \citep{henry00,charbonneau00}, while those systems initially
detected through photometric surveillance require Doppler follow-up
observations to confirm the planet's existence
\citep[e.g.][]{bakos:2007a}.

The combination of Doppler-shift measurements and transits can also
reveal an interesting aspect of the planetary system's architecture:
the orientation of the planet's orbital plane with respect to the
star's spin axis.  This is done by measuring the apparent Doppler shift
of the star throughout a transit.  As the planet's shadow traverses the
rotating stellar surface it alternately blocks the approaching and
receding limbs, giving rise to anomalous Doppler shift known as the
Rossiter-McLaughlin effect \citep{rossiter:1924,mclaughlin:1924}.  By modeling this effect one may determine
the angle $\lambda$ on the sky between the angular momentum vectors of
the planetary orbit and the stellar spin \citep{queloz00, ohta05,
winn05, gaudi07}.

The spin-orbit angles observed among transiting planets have provided
clues about the processes that give rise to the population of close-in
planets.  Giant exoplanets are thought to have formed on wide orbits
and subsequently moved inward to smaller semimajor axes through a
process generically known as orbital migration.  There are many
theories for the mechanisms of inward migration, and each process
leaves behind clues imprinted in the distribution of $\lambda$ observed
among close-in planetary systems.

The initial spin-orbit measurements indicated that the majority of
systems are well-aligned \citep[e.g.][]{winn05}, and this finding
supported the notion that planets likely migrated through a mechanism
such as Type II disk migration that preserves, or even enforces, a
close alignment between the planet orbit normal and stellar spin axis
\citep{winn06,johnson08,fab09}.  However, a larger collection of
spin-orbit measurements has revealed that misaligned, and even
retrograde systems are quite common \citep{hebrard08, winn09,
johnson09, anderson10, triaud10, morton11}.  It now appears that disk
migration is not the only migration channel, and that impulsive,
gravitational interactions are likely responsible for forming many of
the known close-in planetary systems.

\citet{winn10} noted that the misaligned planets tended to be those
having the hottest host stars ($\teffstar > 6250\,K$).  They suggested
that this is a signal that planet migration operates differently for
stars of differing masses.  Another hypothesis put forth by
\citet{winn10} is that cool stars are observed to have low obliquities
only because of tidal interactions between the close-in planet and the
star's relatively thick convective envelope, while hotter stars have
thinner outer envelopes and weaker tidal dissipation.  Indeed, 
\citet{morton11} found that the collection of $\lambda$ measurements
for 12 systems containing hot stars was consistent with all systems
being misaligned\footnote{The measurement of $\lambda$ represents the
  sky-projected spin-orbit angle, not the true 3-dimensional angle
  $\psi$. Thus, the fraction of truly misaligned requires a
  statistical deprojection \citep[e.g.][]{fab09}.}.  If this
interpretation proves to be correct then the 
hot stars are giving a clearer picture of the migration mechanism and
its resulting distribution of angular momentum, while the cooler stars
have been affected by subsequent tidal evolution.

To detect the Rossiter-McLaughlin effect and measure the spin-orbit
angle typically requires a time series of Doppler-shift measurements
with a precision of order 1--10~\ms\ and a time sampling of about 15~min or
better.  While the space-based transit survey mission, \emph{Kepler},
has recently increased the sample of known transiting systems by an
order of magnitude \citep{borucki11}, only the very brightest stars in
the {\it Kepler} sample are amenable to precise measurement of
$\lambda$ \citep[e.g.][]{jenkins10}.  Ground-based transit surveys thus
play a valuable role in understanding exoplanet characteristics by
providing additional, bright ($V < 12$) systems amenable to
high-resolution, spectroscopic follow up.

In this contribution, we announce the discovery of a new transiting
planet orbiting a bright, early-type star, which represents the
thirtieth planet detected by the Hungarian-made Automatic Telescope
Network \citep[HATNet;][]{bakos:2004}. For this system we also
observed the Rossiter-McLaughlin effect, and found that the host star
HAT-P-30 is a member of the growing collection of hot stars with
highly-tilted hot Jupiters.

\begin{figure}[!t]
\plotone{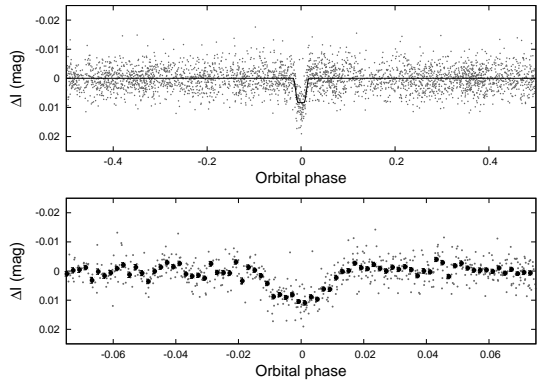}
\caption{
	Unbinned \lc{} of \hatcur{}, including $\approx$3200 data
	points obtained with the HAT-5 and HAT-9 telescopes of HATNet in
	the instrumental \band{I_{C}} with a 5.5-minute cadence (see the
	text for details).  The data have been folded with the period $P =
	\hatcurLCPprec$\,days resulting from the global fit described in
	\refsecl{analysis}).  The solid line shows a simplified transit
	model fit to the light curve (\refsecl{globmod}).  The bottom panel
	shows a zoomed-in view of the transit, the filled circles show the
	light curve binned in phase with a bin-size of 0.002.
    \label{fig:hatnet}}
\end{figure}

\begin{figure} [!ht]
\plotone{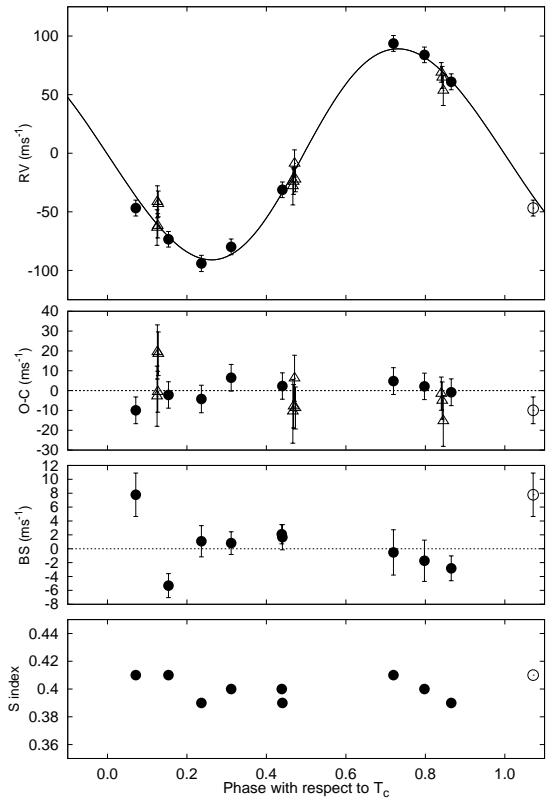}
\caption{
	{\em Top panel:} Keck/HIRES RV measurements (filled circles) and
        Subaru/HDS RV measurements (open triangles) for
        \hbox{\hatcur{}} shown as a function of orbital phase, along
        with our best-fit model (see \reftabl{planetparam}).  Zero
        phase corresponds to the time of mid-transit.  The
        center-of-mass velocity has been subtracted.
	{\em Second panel:} Velocity $O\!-\!C$ residuals from the best
        fit.  The error bars include a component from
        astrophysical/instrumental jitter ($\hatcurRVjitterA$\,\ms\ for
        HIRES and $\hatcurRVjitterB$\,\ms\ for HDS) added in quadrature
        to the formal errors such that $\chi_{\nu}^2 = 1$ (see
        \refsecl{globmod}).
	{\em Third panel:} Bisector spans (BS) for the HIRES spectra, with
        the mean value subtracted, and the measurement from the
        template spectrum is included.
	{\em Bottom panel:} Chromospheric activity index $S$
        measured from the Keck spectra.
	Note the different vertical scales of the panels. Observations
        shown twice are represented with open circles.
\label{fig:rvbis}}
\end{figure}

\ifthenelse{\boolean{emulateapj}}{
    \begin{deluxetable*}{lrrrrrrr}[!ht]
}{
    \begin{deluxetable}{lrrrrrrr}
}
\tabletypesize{\footnotesize}
\tablewidth{0pc}
\tablecaption{
	Relative radial velocities, bisector spans, and activity index
	measurements of \hatcur{}. 
	\label{tab:rvs}
}
\tablehead{
	\colhead{BJD\tablenotemark{a}} & 
	\colhead{RV\tablenotemark{b}} & 
	\colhead{\ensuremath{\sigma_{\rm RV}}\tablenotemark{c}} & 
	\colhead{BS} & 
	\colhead{\ensuremath{\sigma_{\rm BS}}} & 
	\colhead{S\tablenotemark{d}} & 
	\colhead{Phase} &
        \colhead{Instrument}\\
	\colhead{\hbox{(2,454,000$+$)}} & 
	\colhead{(\ms)} & 
	\colhead{(\ms)} &
	\colhead{(\ms)} &
    \colhead{(\ms)} &
	\colhead{} &
	\colhead{} &
        \colhead{}
}
\startdata
\ifthenelse{\boolean{rvtablelong}}{
$ 1313.78961 $ & $   -95.85 $ & $     3.34 $ & $     1.08 $ & $     2.25 $ & $    0.1250 $ & $   0.236 $ &  Keck \\
$ 1320.76920 $ & $    84.53 $ & $     3.11 $ & $    -0.53 $ & $     3.26 $ & $    0.1280 $ & $   0.720 $ &  Keck \\
$ 1321.75751 $ & $   -48.39 $ & $     3.23 $ & $     7.78 $ & $     3.13 $ & $    0.1310 $ & $   0.071 $ &  Keck \\
$ 1338.77142 $ & $   -63.42 $ & $    15.02 $ & \nodata      & \nodata      & \nodata      & $   0.125 $ & Subaru \\
$ 1338.77361 $ & $   -41.42 $ & $    13.48 $ & \nodata      & \nodata      & \nodata      & $   0.126 $ & Subaru \\
$ 1338.77647 $ & $   -61.91 $ & $    10.11 $ & \nodata      & \nodata      & \nodata      & $   0.127 $ & Subaru \\
$ 1338.78003 $ & $   -43.23 $ & $    10.80 $ & \nodata      & \nodata      & \nodata      & $   0.128 $ & Subaru \\
$ 1339.73204 $ & $   -28.01 $ & $    15.95 $ & \nodata      & \nodata      & \nodata      & $   0.467 $ & Subaru \\
$ 1339.73757 $ & $   -24.22 $ & $    10.72 $ & \nodata      & \nodata      & \nodata      & $   0.469 $ & Subaru \\
$ 1339.74390 $ & $    -8.72 $ & $    11.41 $ & \nodata      & \nodata      & \nodata      & $   0.471 $ & Subaru \\
$ 1339.75024 $ & $   -22.31 $ & $    10.37 $ & \nodata      & \nodata      & \nodata      & $   0.473 $ & Subaru \\
$ 1340.78173 $ & $    68.96 $ & $     8.10 $ & \nodata      & \nodata      & \nodata      & $   0.840 $ & Subaru \\
$ 1340.78830 $ & $    64.70 $ & $     9.16 $ & \nodata      & \nodata      & \nodata      & $   0.842 $ & Subaru \\
$ 1340.79604 $ & $    53.51 $ & $    12.62 $ & \nodata      & \nodata      & \nodata      & $   0.845 $ & Subaru \\
$ 1466.12996 $ & \nodata      & \nodata      & $     2.09 $ & $     1.39 $ & $    0.1280 $ & $   0.439 $ &  Keck \\
$ 1466.13428 $ & $   -34.89 $ & $     3.15 $ & $     1.66 $ & $     1.81 $ & $    0.1270 $ & $   0.440 $ &  Keck \\
$ 1467.13927 $ & $    84.32 $ & $     2.78 $ & $    -1.73 $ & $     2.97 $ & $    0.1290 $ & $   0.798 $ &  Keck \\
$ 1468.13943 $ & $   -74.68 $ & $     2.79 $ & $    -5.31 $ & $     1.73 $ & $    0.1300 $ & $   0.154 $ &  Keck \\
$ 1470.13939 $ & $    59.23 $ & $     3.06 $ & $    -2.82 $ & $     1.79 $ & $    0.1250 $ & $   0.865 $ &  Keck \\
$ 1491.06766 $ & $   -85.18 $ & $     2.83 $ & $     0.81 $ & $     1.64 $ & $    0.1290 $ & $   0.311 $ &  Keck \\
$ 1613.74484 $ & $    15.56 $ & $     2.63 $ & \nodata      & \nodata      & \nodata      & $   0.959 $ & Keck \\
$ 1613.74950 $ & $    11.06 $ & $     2.54 $ & \nodata      & \nodata      & \nodata      & $   0.961 $ & Keck \\
$ 1613.75421 $ & $    24.38 $ & $     2.59 $ & \nodata      & \nodata      & \nodata      & $   0.963 $ & Keck \\
$ 1613.76193 $ & $    19.07 $ & $     2.95 $ & \nodata      & \nodata      & \nodata      & $   0.966 $ & Keck \\
$ 1613.77038 $ & $    10.54 $ & $     2.45 $ & \nodata      & \nodata      & \nodata      & $   0.969 $ & Keck \\
$ 1613.77489 $ & $     8.84 $ & $     2.59 $ & \nodata      & \nodata      & \nodata      & $   0.970 $ & Keck \\
$ 1613.77909 $ & $    12.27 $ & $     2.47 $ & \nodata      & \nodata      & \nodata      & $   0.972 $ & Keck \\
$ 1613.78283 $ & $    16.46 $ & $     2.59 $ & \nodata      & \nodata      & \nodata      & $   0.973 $ & Keck \\
$ 1613.78655 $ & $    15.22 $ & $     2.47 $ & \nodata      & \nodata      & \nodata      & $   0.974 $ & Keck \\
$ 1613.79003 $ & $    14.57 $ & $     2.42 $ & \nodata      & \nodata      & \nodata      & $   0.976 $ & Keck \\
$ 1613.79344 $ & $    16.91 $ & $     2.73 $ & \nodata      & \nodata      & \nodata      & $   0.977 $ & Keck \\
$ 1613.79688 $ & $    18.70 $ & $     2.58 $ & \nodata      & \nodata      & \nodata      & $   0.978 $ & Keck \\
$ 1613.80058 $ & $    10.80 $ & $     2.47 $ & \nodata      & \nodata      & \nodata      & $   0.979 $ & Keck \\
$ 1613.80414 $ & $    11.91 $ & $     2.43 $ & \nodata      & \nodata      & \nodata      & $   0.981 $ & Keck \\
$ 1613.80765 $ & $     9.29 $ & $     2.40 $ & \nodata      & \nodata      & \nodata      & $   0.982 $ & Keck \\
$ 1613.81152 $ & $    -3.95 $ & $     2.41 $ & \nodata      & \nodata      & \nodata      & $   0.983 $ & Keck \\
$ 1613.81606 $ & $    10.13 $ & $     2.40 $ & \nodata      & \nodata      & \nodata      & $   0.985 $ & Keck \\
$ 1613.81974 $ & $     5.64 $ & $     2.57 $ & \nodata      & \nodata      & \nodata      & $   0.986 $ & Keck \\
$ 1613.82300 $ & $     4.62 $ & $     2.59 $ & \nodata      & \nodata      & \nodata      & $   0.987 $ & Keck \\
$ 1613.82623 $ & $     6.70 $ & $     2.29 $ & \nodata      & \nodata      & \nodata      & $   0.988 $ & Keck \\
$ 1613.82951 $ & $   -10.97 $ & $     2.41 $ & \nodata      & \nodata      & \nodata      & $   0.990 $ & Keck \\
$ 1613.83289 $ & $    -9.04 $ & $     2.46 $ & \nodata      & \nodata      & \nodata      & $   0.991 $ & Keck \\
$ 1613.83642 $ & $   -19.67 $ & $     2.24 $ & \nodata      & \nodata      & \nodata      & $   0.992 $ & Keck \\
$ 1613.83980 $ & $   -29.85 $ & $     2.51 $ & \nodata      & \nodata      & \nodata      & $   0.993 $ & Keck \\
$ 1613.84303 $ & $   -13.80 $ & $     2.57 $ & \nodata      & \nodata      & \nodata      & $   0.994 $ & Keck \\
$ 1613.84623 $ & $   -23.45 $ & $     2.33 $ & \nodata      & \nodata      & \nodata      & $   0.996 $ & Keck \\
$ 1613.84972 $ & $   -23.52 $ & $     2.77 $ & \nodata      & \nodata      & \nodata      & $   0.997 $ & Keck \\
$ 1613.85375 $ & $   -29.80 $ & $     2.73 $ & \nodata      & \nodata      & \nodata      & $   0.998 $ & Keck \\
$ 1613.85753 $ & $   -35.90 $ & $     2.53 $ & \nodata      & \nodata      & \nodata      & $   1.000 $ & Keck \\
$ 1613.86098 $ & $   -27.89 $ & $     2.37 $ & \nodata      & \nodata      & \nodata      & $   0.001 $ & Keck \\
$ 1613.86452 $ & $   -28.97 $ & $     2.18 $ & \nodata      & \nodata      & \nodata      & $   0.002 $ & Keck \\
$ 1613.86772 $ & $   -31.21 $ & $     2.51 $ & \nodata      & \nodata      & \nodata      & $   0.003 $ & Keck \\
$ 1613.87093 $ & $   -34.59 $ & $     2.37 $ & \nodata      & \nodata      & \nodata      & $   0.004 $ & Keck \\
$ 1613.87414 $ & $   -33.40 $ & $     2.33 $ & \nodata      & \nodata      & \nodata      & $   0.005 $ & Keck \\
$ 1613.87738 $ & $   -23.28 $ & $     2.32 $ & \nodata      & \nodata      & \nodata      & $   0.007 $ & Keck \\
$ 1613.88045 $ & $   -32.05 $ & $     2.29 $ & \nodata      & \nodata      & \nodata      & $   0.008 $ & Keck \\
$ 1613.88359 $ & $   -30.53 $ & $     2.18 $ & \nodata      & \nodata      & \nodata      & $   0.009 $ & Keck \\
$ 1613.88694 $ & $   -28.82 $ & $     2.42 $ & \nodata      & \nodata      & \nodata      & $   0.010 $ & Keck \\
$ 1613.89023 $ & $   -23.92 $ & $     2.40 $ & \nodata      & \nodata      & \nodata      & $   0.011 $ & Keck \\
$ 1613.89374 $ & $   -28.03 $ & $     2.42 $ & \nodata      & \nodata      & \nodata      & $   0.012 $ & Keck \\
$ 1613.89709 $ & $   -15.26 $ & $     2.42 $ & \nodata      & \nodata      & \nodata      & $   0.014 $ & Keck \\
$ 1613.90041 $ & $    -8.88 $ & $     2.56 $ & \nodata      & \nodata      & \nodata      & $   0.015 $ & Keck \\
$ 1613.90369 $ & $   -14.35 $ & $     2.49 $ & \nodata      & \nodata      & \nodata      & $   0.016 $ & Keck \\
$ 1613.90688 $ & $   -14.53 $ & $     2.41 $ & \nodata      & \nodata      & \nodata      & $   0.017 $ & Keck \\
$ 1613.91005 $ & $    -1.77 $ & $     2.38 $ & \nodata      & \nodata      & \nodata      & $   0.018 $ & Keck \\
$ 1613.91328 $ & $   -15.95 $ & $     2.45 $ & \nodata      & \nodata      & \nodata      & $   0.019 $ & Keck \\
$ 1613.91645 $ & $   -14.47 $ & $     2.24 $ & \nodata      & \nodata      & \nodata      & $   0.021 $ & Keck \\
$ 1613.91968 $ & $    -5.13 $ & $     2.30 $ & \nodata      & \nodata      & \nodata      & $   0.022 $ & Keck \\
$ 1613.92286 $ & $   -10.78 $ & $     2.45 $ & \nodata      & \nodata      & \nodata      & $   0.023 $ & Keck \\
$ 1613.92602 $ & $    -8.85 $ & $     2.45 $ & \nodata      & \nodata      & \nodata      & $   0.024 $ & Keck \\
$ 1613.92912 $ & $   -12.73 $ & $     2.54 $ & \nodata      & \nodata      & \nodata      & $   0.025 $ & Keck \\
$ 1613.93228 $ & $   -22.34 $ & $     2.39 $ & \nodata      & \nodata      & \nodata      & $   0.026 $ & Keck \\
$ 1613.93550 $ & $   -19.25 $ & $     2.28 $ & \nodata      & \nodata      & \nodata      & $   0.027 $ & Keck \\
$ 1613.93874 $ & $   -16.53 $ & $     2.54 $ & \nodata      & \nodata      & \nodata      & $   0.028 $ & Keck \\
$ 1613.94200 $ & $   -16.67 $ & $     2.45 $ & \nodata      & \nodata      & \nodata      & $   0.030 $ & Keck \\
$ 1614.00550 $ & $   -38.59 $ & $     2.59 $ & \nodata      & \nodata      & \nodata      & $   0.052 $ & Keck \\
$ 1614.00893 $ & $   -36.68 $ & $     2.66 $ & \nodata      & \nodata      & \nodata      & $   0.053 $ & Keck \\
$ 1614.01252 $ & $   -31.57 $ & $     2.38 $ & \nodata      & \nodata      & \nodata      & $   0.055 $ & Keck \\
    
}{
$ 1313.78961 $ & $   -95.85 $ & $     3.34 $ & $     1.08 $ & $     2.25 $ & $    0.1250 $ & $   0.236 $ &  Keck \\
$ 1320.76920 $ & $    84.53 $ & $     3.11 $ & $    -0.53 $ & $     3.26 $ & $    0.1280 $ & $   0.720 $ &  Keck \\
$ 1321.75751 $ & $   -48.39 $ & $     3.23 $ & $     7.78 $ & $     3.13 $ & $    0.1310 $ & $   0.071 $ &  Keck \\
$ 1338.77142 $ & $   -63.42 $ & $    15.02 $ & \nodata      & \nodata      & \nodata      & $   0.125 $ & Subaru \\
$ 1338.77361 $ & $   -41.42 $ & $    13.48 $ & \nodata      & \nodata      & \nodata      & $   0.126 $ & Subaru \\
$ 1338.77647 $ & $   -61.91 $ & $    10.11 $ & \nodata      & \nodata      & \nodata      & $   0.127 $ & Subaru \\
$ 1338.78003 $ & $   -43.23 $ & $    10.80 $ & \nodata      & \nodata      & \nodata      & $   0.128 $ & Subaru \\
$ 1339.73204 $ & $   -28.01 $ & $    15.95 $ & \nodata      & \nodata      & \nodata      & $   0.467 $ & Subaru \\
$ 1339.73757 $ & $   -24.22 $ & $    10.72 $ & \nodata      & \nodata      & \nodata      & $   0.469 $ & Subaru \\
$ 1339.74390 $ & $    -8.72 $ & $    11.41 $ & \nodata      & \nodata      & \nodata      & $   0.471 $ & Subaru \\
$ 1339.75024 $ & $   -22.31 $ & $    10.37 $ & \nodata      & \nodata      & \nodata      & $   0.473 $ & Subaru \\
    
}
\enddata
\tablenotetext{a}{
    Barycentric Julian dates throughout the
    paper are calculated from Coordinated Universal Time (UTC).
}
\tablenotetext{b}{
	The zero-point of these velocities is arbitrary. An overall offset
    $\gamma_{\rm rel}$ fitted to these velocities in \refsecl{globmod}
    has {\em not} been subtracted.
}
\tablenotetext{c}{
	Internal errors excluding the component of astrophysical/instrumental jitter
    considered in \refsecl{globmod}.
}
\tablenotetext{d}{
	Chromospheric activity index, calibrated to the
	scale of Isaacson \& Fischer (2010).
}
\ifthenelse{\boolean{rvtablelong}}{
	\tablecomments{
		 For the iodine-free template exposures
                there is no RV 
		measurement, but the BS and S index can still be determined.
	}
}{
    \tablecomments{
		For the iodine-free template exposures there is no RV
		measurement, but the BS and S index can still be determined.
		This table is presented in its entirety in the
		electronic edition of the Astrophysical Journal.  A portion is
		shown here for guidance regarding its form and content.
	}
} 
\ifthenelse{\boolean{emulateapj}}{
    \end{deluxetable*}
}{
    \end{deluxetable}
}

\section{Observations}
\label{sec:obs}

The HATNet observing strategy has been described in detail in previous
articles \citep[e.g.][]{bakos:2010,latham:2009}, and we summarize it
briefly as follows.  Photometric observations with one or more of the
HATNet telescopes initially identify stars exhibiting periodic dimming
events that resemble the signals of planetary transits.  These
candidates are then followed up using high-resolution, low-S/N
``reconnaissance'' spectroscopic observations using 1-2\,m class
telescopes.  The spectroscopic observations allow many false positives
to be rejected (e.g.\ unresolved blends of bright stars with background
eclipsing binaries).  Additional transit light curves are acquired to
refine the light curve properties.  Finally, high-resolution, high-S/N
``characterization'' spectroscopy is undertaken, with the goals of
detecting the orbital motion of the star due to the planet,
characterizing the host star, and ruling out subtle blend scenarios by
detecting line bisector variations.  In the following subsections we
highlight specific details of this procedure that are pertinent to the
discovery of \hatcurb{}.

\subsection{Photometric detection}
\label{sec:detection}

The transits of \hatcur{} ($=$\hatcurCCgsc{}) were detected with the
HAT-5 telescope in Arizona and the HAT-9 telescope in Hawaii, within a
target field internally labeled as \hatcurfield.  The field was
observed nightly between 2008 December and 2009 May, whenever weather
conditions permitted.  We gathered 3686 images, each with an exposure
time of 5 minutes, and an observing cadence of 5.5~minutes. 
Approximately 500 images were rejected by our reduction pipeline
because they were of relatively poor quality.  Each image encompasses
about 66,000 stars with $r$ magnitudes brighter than 14.5.  For the
brightest stars in the field, we achieved a per-image photometric
precision of 4\,mmag.

The HATNet images were calibrated, and trend-filtered light curves were
derived for stars in the Two Micron All Sky Survey catalog
\citep[2MASS;][]{skrutskie:2006} following the procedure described by
\citet{pal:2006,pal:2009b}.  We searched the photometric time series of
each target using the Box Least-Squares \citep[BLS;][]{kovacs:2002}
method.  We detected a significant signal in the \lc{} of
\hatcurCCgsc{} ($\alpha = \hatcurCCra$, $\delta = \hatcurCCdec$; J2000;
V=\hatcurCCtassmv\, \citealp{droege:2006}), with an apparent depth of
$\sim\hatcurLCdip$\,mmag, and a period of $P=\hatcurLCPshort$\,days
(see \reffigl{hatnet}).  This star is henceforth referred to as
\hatcur{}.

\subsection{Reconnaissance Spectroscopy}
\label{sec:recspec}

Four spectra of \hatcur\ were obtained with the Tillinghast Reflector
Echelle Spectrograph \citep{furesz:08} in March 2010, for an initial
reconnaissance of the candidate, and thus were not particularly strong
exposures; the typical SNR per resolution element was 50 near the Mg b
features.  Classification of these spectra using techniques similar to
those described by \citet{latham:2009} showed that the star is a
slowly-rotating early G dwarf with sharp lines suitable for very precise
radial velocities.  Multi-order relative radial velocities were derived
using the techniques described by \citet{buchhave:2010}.  A
template for the cross-correlations analysis was created by shifting
and co-adding the
four individual observations to a common velocity.

A circular orbit with the period and phase set by the photometric ephemeris
yielded a good fit with semi-amplitude $K = 84.5 \pm 9.8$ \ms.  This was
interpreted as strong evidence that the companion is a Hot Jupiter, and
therefore \hatcur\ was scheduled for additional high-resolution
spectroscopic observations with larget telescopes to provide a
higher-quality orbital solution and more precise mass determination,
as described in \S~\ref{sec:hispec}.

\subsection{High resolution, high S/N spectroscopy}
\label{sec:hispec}

We proceeded with the follow-up of this candidate by obtaining
high-resolution, high-S/N spectra to measure the spectroscopic orbit of
the system, and refine the physical parameters of the host star. 
For this we used the High-Resolution Echelle Spectrometer on the Keck~I
telescope \citep[HIRES;][]{vogt:1994}, and the High-Dispersion
Spectrograph \citep[HDS;][]{Noguchi:2002} on the Subaru telescope, both
atop Mauna Kea in Hawaii.  For both telescope/spectrometer
configurations we used the iodine cell method, which has been described
by \citet{marcy:1992} and \citet{butler:1996}.  The Keck
implementation of this method has been 
described by \citet{howard10} and \citet{johnson09}, and the Subaru
implementation has been described by \citet{sato02,sato05}.  The
resulting RV measurements and their uncertainties are listed in
\reftabl{rvs}, and the phase-folded RV measurements and best fitting
orbit are shown in \reffigl{rvbis} (see also \refsecl{analysis}).

We also checked if the measured radial velocities are not real, but are
instead caused by distortions in the spectral line profiles due to
contamination from an unresolved eclipsing binary
\citep[e.g.][]{torres:2007}.  We performed a bisector analysis on the
Keck spectra using the method described in \S~5 of \cite{bakos:2007a}. 
The resulting bisector spans show no significant variation and are not
correlated with the RVs (\reffigl{rvbis}), which argues against
false-positive  scenarios involving a blended, background eclipsing
binary system. 

There is no sign of emission in the cores of the \ion{Ca}{2} H and K
lines in any of our spectra, from which we conclude that the
chromospheric activity level in \hatcur{} is very low and that the star
is fairly old ($ > 1$~Gyr).  In \reffigl{rvbis} we also show the $S$
index \citep{vaughan:1978, isaacson10}, which is a measure of the
chromospheric activity of the star derived from the cores of the
\ion{Ca}{2} H and K lines.  We do not detect significant variation of
the index as a function of orbital phase.  It is therefore unlikely
that the observed RV variations are due to stellar activity.

\subsection{Photometric follow-up observations}
\label{sec:phot}

\begin{figure}[!t]
\plotone{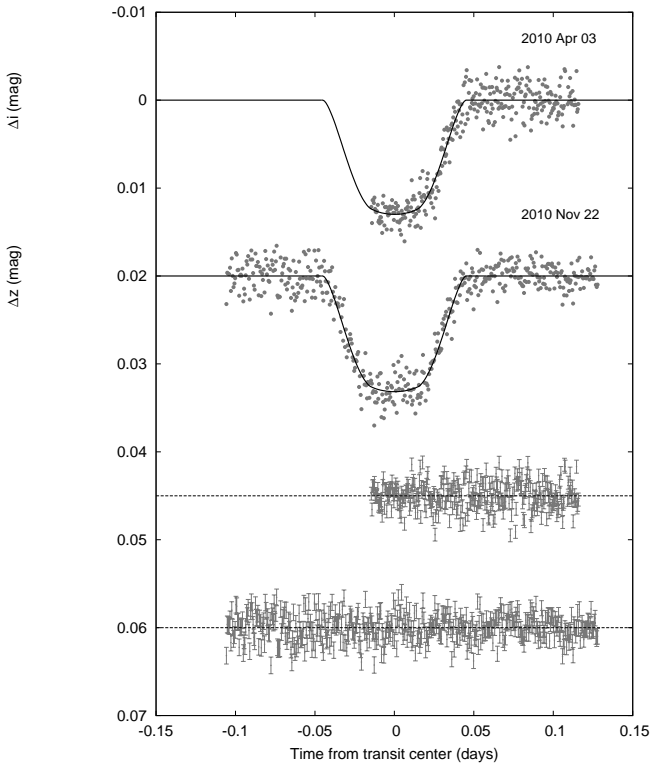}
\caption{
	Unbinned instrumental \band{i} and \band{z} transit \lcs{},
	acquired with KeplerCam at the \flwof{} telescope.  The light
	curves are displaced vertically for clarity.  Our best fit from the
	global modeling described in \refsecl{globmod} is shown by the
	solid lines.  Residuals from the fits are displayed at the bottom,
	in the same order as the top curves.  \label{fig:lc}}
\end{figure}

We acquired additional light curves using the KeplerCam imager on the
1.2~m telescope at the Fred Lawrence Whipple Observatory (FLWO).  We
observed a partial transit on UT 2010 April 3 through the a Sloan
\band{i} filter, and a full transit on UT 2010 November 22 using a
\band{z} filter.  The reduction of the images, including basic
calibration, astrometry, and aperture photometry, was performed as
described by \citet{bakos:2010} The final time series are shown in the
top portion of \reffigl{lc}, along with our best-fit transit \lc{}
model described below; the individual measurements are reported in
\reftabl{phfu}.

\begin{deluxetable}{lrrrr}
\tablewidth{0pc}
\tablecaption{KeplerCam differential photometry of 
	\hatcur\label{tab:phfu}}
\tablehead{
	\colhead{BJD} & 
	\colhead{Mag\tablenotemark{a}} & 
	\colhead{\ensuremath{\sigma_{\rm Mag}}} &
	\colhead{Mag(orig)\tablenotemark{b}} & 
	\colhead{Filter} \\
	\colhead{\hbox{~~~~(2,400,000$+$)~~~~}} & 
	\colhead{} & 
	\colhead{} &
	\colhead{} & 
	\colhead{}
}
\startdata
$ 55290.62596 $ & $   0.01214 $ & $   0.00073 $ & $   9.28160 $ & $ i$\\
$ 55290.62636 $ & $   0.01398 $ & $   0.00073 $ & $   9.28349 $ & $ i$\\
$ 55290.62675 $ & $   0.01138 $ & $   0.00073 $ & $   9.28059 $ & $ i$\\
$ 55290.62715 $ & $   0.01136 $ & $   0.00073 $ & $   9.28017 $ & $ i$\\
$ 55290.62754 $ & $   0.01169 $ & $   0.00073 $ & $   9.28064 $ & $ i$\\
$ 55290.62793 $ & $   0.01108 $ & $   0.00073 $ & $   9.28062 $ & $ i$\\
$ 55290.62833 $ & $   0.01237 $ & $   0.00073 $ & $   9.28116 $ & $ i$\\
$ 55290.62871 $ & $   0.01269 $ & $   0.00073 $ & $   9.28152 $ & $ i$\\
$ 55290.62911 $ & $   0.01274 $ & $   0.00073 $ & $   9.28093 $ & $ i$\\
$ 55290.62948 $ & $   0.01261 $ & $   0.00073 $ & $   9.28249 $ & $ i$\\

[-1.5ex]
\enddata
\tablenotetext{a}{
	The out-of-transit level has been subtracted. These magnitudes have
	been subjected to the EPD and TFA procedures.
}
\tablenotetext{b}{
	Raw magnitude values without application of the EPD and TFA
	procedures.
}
\tablecomments{
    This table is available in a machine-readable form in the online
    journal.  A portion is shown here for guidance regarding its form
    and content.
}
\end{deluxetable}

\section{Analysis}
\label{sec:analysis}

The analysis of the \hatcur{} system, including determinations of the
properties of the host star and planet, was carried out in a similar
fashion to previous HATNet discoveries \citep[e.g.][]{bakos:2010}. 
Below we briefly summarize the procedure and the results for the
\hatcurb{} system.

\subsection{Properties of the parent star}
\label{sec:stelparam}

We estimate stellar atmospheric parameters based on the iodine-free
``template'' spectrum of the star obtained with the Keck/HIRES
instrument.  We fitted the spectra using synthetic spectra generated by
the LTE analysis package known as Spectroscopy Made Easy
\citep[SME;][]{valenti:1996}, with the atomic line database of
\cite{valenti:2005}.  SME provides the following initial values and
uncertainties:
effective temperature $\teffstar=\hatcurSMEiteff$\,K, metallicity
$\feh=\hatcurSMEizfeh$\,dex, and stellar surface gravity
$\loggstar=\hatcurSMEilogg$\,(cgs), projected rotational velocity
$\vsini=\hatcurSMEivsin\,\kms$.

We use these parameters to determine the limb-darkening coefficients
needed in the global modeling of the follow-up photometry.  This
modeling strongly constrains the stellar density \rhostar, which
combined with the spectroscopically determined \teffstar\ and \feh\ and
the Yonsei-Yale stellar evolution models \citep[YY;][]{yi:2001}
provides a refined estimate of \loggstar.  We perform a second
iteration of SME with \loggstar\ fixed to this value, followed by a
second iteration of the global modeling and comparison to the YY
models.  The resulting \loggstar\ was consistent with the previous
value so that no further iterations of this procedure were needed.  Our
final adopted stellar parameters are listed in \reftabl{stellar}.  We
find that \hatcur\ is a Gyr-old F-type dwarf star with a mass of
$\mstar = \hatcurISOm\msun$ and radius of $\rstar = \hatcurISOr\rsun$.

\begin{deluxetable}{lrl}
\tablewidth{0pc}
\tabletypesize{\scriptsize}
\tablecaption{
	Stellar parameters for \hatcur{}
	\label{tab:stellar}
}
\tablehead{
	\colhead{~~~~~~~~Parameter~~~~~~~~}	&
	\colhead{Value} &
	\colhead{Source}
}
\startdata
\noalign{\vskip -3pt}
\sidehead{Spectroscopic properties}
~~~~$\teffstar$ (K)\dotfill         &  \hatcurSMEteff       & SME\tablenotemark{a}\\
~~~~$\feh$\dotfill                  &  \hatcurSMEzfeh       & SME                 \\
~~~~$\vsini$ (\kms)\dotfill         &  \hatcurSMEvsin       & SME                 \\
~~~~$\vmac$ (\kms)\dotfill          &  \hatcurSMEvmac       & SME                 \\
~~~~$\vmic$ (\kms)\dotfill          &  \hatcurSMEvmic       & SME                 \\
~~~~$\gamma_{\rm RV}$ (\kms)\dotfill&  \hatcurTRESgamma     & TRES                  \\
\sidehead{Photometric properties}
~~~~$V$ (mag)\dotfill               &  \hatcurCCtassmv      & TASS                \\
~~~~$V\!-\!I_C$ (mag)\dotfill       &  \hatcurCCtassvi      & TASS                \\
~~~~$J$ (mag)\dotfill               &  \hatcurCCtwomassJmag & 2MASS           \\
~~~~$H$ (mag)\dotfill               &  \hatcurCCtwomassHmag & 2MASS           \\
~~~~$K_s$ (mag)\dotfill             &  \hatcurCCtwomassKmag & 2MASS           \\
\sidehead{Derived properties}
~~~~$\mstar$ ($\msun$)\dotfill      &  \hatcurISOmlong      & \hatcurisoshort+\hatcurlumind+SME \tablenotemark{b}\\
~~~~$\rstar$ ($\rsun$)\dotfill      &  \hatcurISOrlong      & \hatcurisoshort+\hatcurlumind+SME         \\
~~~~$\loggstar$ (cgs)\dotfill       &  \hatcurISOlogg       & \hatcurisoshort+\hatcurlumind+SME         \\
~~~~$\lstar$ ($\lsun$)\dotfill      &  \hatcurISOlum        & \hatcurisoshort+\hatcurlumind+SME         \\
~~~~$M_V$ (mag)\dotfill             &  \hatcurISOmv         & \hatcurisoshort+\hatcurlumind+SME         \\
~~~~$M_K$ (mag,\hatcurjhkfilset)\dotfill &  \hatcurISOMK    & \hatcurisoshort+\hatcurlumind+SME         \\
~~~~Age (Gyr)\dotfill               &  \hatcurISOage        & \hatcurisoshort+\hatcurlumind+SME         \\
~~~~Distance (pc)\dotfill           &  \hatcurXdist         & \hatcurisoshort+\hatcurlumind+SME+2MASS\\
[-1.5ex]
\enddata
\tablenotetext{a}{
	SME = ``Spectroscopy Made Easy'' package for the analysis of
	high-resolution spectra \citep{valenti:1996}.  These parameters
	rely primarily on SME, but have a small dependence also on the
	iterative analysis incorporating the isochrone search and global
	modeling of the data, as described in the text.
}
\tablenotetext{b}{
	\hatcurisoshort+\hatcurlumind+SME = Based on the \hatcurisoshort\
    isochrones \citep{\hatcurisocite}, \hatcurlumind\ as a luminosity
    indicator, and the SME results.
}
\end{deluxetable}

\subsection{Global modeling of the data}
\label{sec:globmod}

We modeled the HATNet photometry and follow-up RV measurements using
the procedure described in detail by \citet{bakos:2010}.  The resulting
parameters pertaining to the light curves and RV curves, together with
derived physical parameters of the planet, are listed in
\reftabl{planetparam}. 

Based on the amplitude of the RV variations, together with the stellar
mass and its associated uncertainty, we estimate a mass for the planet
of
$\mpl=\hatcurPPmlong\,\mjup$. The transit parameters give a planetary 
radius of
$\rpl=\hatcurPPrlong\,\rjup$, leading to a mean planetary density
$\rho_p=\hatcurPPrho$\,\gcmc. 
These and other planetary parameters are listed at the bottom of
Table~\ref{tab:planetparam}. We note that the eccentricity of the
orbit is consistent with zero: $e = \hatcurRVeccen$, $\omega =
\hatcurRVomega\arcdeg$.

\begin{deluxetable}{lr}
\tabletypesize{\scriptsize}
\tablecaption{Orbital and planetary parameters\label{tab:planetparam}}
\tablehead{
	\colhead{~~~~~~~~~~~~~~~Parameter~~~~~~~~~~~~~~~} &
	\colhead{Value}
}
\startdata
\noalign{\vskip -3pt}
\sidehead{\Lc{} parameters}
~~~$P$ (days)             \dotfill    & $\hatcurLCP$              \\
~~~$T_c$ (${\rm BJD}$)    
      \tablenotemark{a}   \dotfill    & $\hatcurLCT$              \\
~~~$T_{14}$ (days)
      \tablenotemark{a}   \dotfill    & $\hatcurLCdur$            \\
~~~$T_{12} = T_{34}$ (days)
      \tablenotemark{a}   \dotfill    & $\hatcurLCingdur$         \\
~~~$\arstar$              \dotfill    & $\hatcurPPar$             \\
~~~$\zrstar$              \dotfill    & $\hatcurLCzeta$           \\
~~~$\rpl/\rstar$          \dotfill    & $\hatcurLCrprstar$        \\
~~~$b^2$                  \dotfill    & $\hatcurLCbsq$            \\
~~~$b \equiv a \cos i/\rstar$
                          \dotfill    & $\hatcurLCimp$            \\
~~~$i$ (deg)              \dotfill    & $\hatcurPPi$              \\

\sidehead{Limb-darkening coefficients \tablenotemark{b}}
~~~$a_i$ (linear term, $i$ filter)    \dotfill    & $\hatcurLBii$             \\
~~~$b_i$ (quadratic term) \dotfill    & $\hatcurLBiii$            \\
~~~$a_z$                 \dotfill    & $\hatcurLBiz$             \\
~~~$b_z$                 \dotfill    & $\hatcurLBiiz$            \\

\sidehead{RV parameters}
~~~$K$ (\ms)              \dotfill    & $\hatcurRVK$              \\
~~~$k_{\rm RV}$\tablenotemark{c} 
                          \dotfill    & $\hatcurRVk$              \\
~~~$h_{\rm RV}$\tablenotemark{c}
                          \dotfill    & $\hatcurRVh$              \\
~~~$e$                    \dotfill    & $\hatcurRVeccen$          \\
~~~$\omega$ (deg)         \dotfill    & $\hatcurRVomega$          \\
~~~HIRES RV fit rms (\ms)        \dotfill    & \hatcurRVfitrmsA           \\
~~~HDS RV fit rms (\ms)        \dotfill    & \hatcurRVfitrmsB           \\

\sidehead{Secondary eclipse parameters}
~~~$T_s$ (BJD)            \dotfill    & $\hatcurXsecondary$       \\
~~~$T_{s,14}$             \dotfill    & $\hatcurXsecdur$          \\
~~~$T_{s,12}$             \dotfill    & $\hatcurXsecingdur$       \\

\sidehead{Planetary parameters}
~~~$\mpl$ ($\mjup$)       \dotfill    & $\hatcurPPmlong$          \\
~~~$\rpl$ ($\rjup$)       \dotfill    & $\hatcurPPrlong$          \\
~~~$C(\mpl,\rpl)$
    \tablenotemark{d}     \dotfill    & $\hatcurPPmrcorr$         \\
~~~$\rhopl$ (\gcmc)       \dotfill    & $\hatcurPPrho$            \\
~~~$\log g_p$ (cgs)       \dotfill    & $\hatcurPPlogg$           \\
~~~$a$ (AU)               \dotfill    & $\hatcurPParel$           \\
~~~$T_{\rm eq}$ (K)       \dotfill    & $\hatcurPPteff$           \\
~~~$\Theta$\tablenotemark{e}\dotfill  & $\hatcurPPtheta$          \\
~~~$F_{per}$ ($10^{\hatcurPPfluxperidim}$\ergscmsq) \tablenotemark{f}
                          \dotfill    & $\hatcurPPfluxperi$       \\
~~~$F_{ap}$  ($10^{\hatcurPPfluxapdim}$\ergscmsq) \tablenotemark{f} 
                          \dotfill    & $\hatcurPPfluxap$         \\
~~~$\langle F \rangle$ ($10^{\hatcurPPfluxavgdim}$\ergscmsq) 
\tablenotemark{f}         \dotfill    & $\hatcurPPfluxavg$        \\
[-1.5ex]

\sidehead{Parameters from Rossiter-McLaughlin Effect}

~~~Projected spin-orbit angle, $\lambda$~[deg] \dotfill & $73.5 \pm 9.0$ \\
~~~$v \sin i_\star$~[km~s$^{-1}$] \dotfill    & $3.07\pm 0.24$
\enddata
\tablenotetext{a}{
    \ensuremath{T_c}: Reference epoch of mid transit that minimizes the
    correlation with the orbital period.  
    \ensuremath{T_{14}}: total transit duration, time between first to
    last contact;
    \ensuremath{T_{12}=T_{34}}: ingress/egress time, time between first
    and second, or third and fourth contact.
}
\tablenotetext{b}{
	Values for a quadratic law, adopted from the tabulations by
    \cite{claret:2004} according to the spectroscopic (SME) parameters
    listed in \reftabl{stellar}.
}
\tablenotetext{c}{
    Lagrangian orbital parameters derived from the global modeling, and
    primarily determined by the RV data.
}
\tablenotetext{d}{
	Correlation coefficient between the planetary mass \mpl\ and radius
	\rpl.
}
\tablenotetext{e}{
	The Safronov number is given by $\Theta = \frac{1}{2}(V_{\rm
	esc}/V_{\rm orb})^2 = (a/\rpl)(\mpl / \mstar )$
	\citep[see][]{hansen:2007}.
}
\tablenotetext{e}{
	Incoming flux per unit surface area, averaged over the orbit.
}
\end{deluxetable}

\subsection{The Rossiter-McLaughlin effect}

We undertook a separate analysis of the RVs obtained on the transit
night (2011~Feb~21) in order to determine the projected spin-orbit
angle $\lambda$.  Our model for the RV data was the sum of the orbital
radial velocity, the anomalous velocity due to the Rossiter-McLaughlin
effect, and a constant offset:
\begin{equation}
V_{{\rm calc}}(t) = V_{\rm orb}(t) + V_{\rm RM}(t) + \gamma.
\end{equation}
For modeling the RM effect we used the technique of \citet{winn05},
in which simulated spectra exhibiting the RM effect are created and
then analyzed using the Doppler-measurement code.  The resulting
formula for the anomalous velocity was
\begin{equation}
V_{\rm RM}(t) = \Delta f(t)~v_p(t)
\left[1.005 - 0.1141\left(\frac{v_p(t)}{2.2~{\rm km~s}^{-1}}\right)^2\right],
\end{equation}
where $\Delta f$ is the fraction of light blocked by the planet, and
$v_p$ is the projected rotation velocity of the portion of the star
that is hidden by the planet.  In calculating $\Delta f(t)$, we adopted
a linear limb-darkening law.  In calculating $v_p(t)$, we assumed
uniform rotation around an axis that is inclined by an angle $\lambda$
from the orbit normal as projected on the sky (using the same
coordinate system as \citet{ohta05} and \citet{fab09}).

Most of the orbital and transit parameters are much more tightly
constrained by other observations.  For this reason we adopted Gaussian
priors on $R_p/R_\star$, $T_{12}$, $T_{14}$, $T_c$, $P$, $K$, and
$v\sin i_\star$, based on the mean parameter values and 1-$\sigma$
uncertainties quoted in Tables~\ref{tab:stellar} 
and~\ref{tab:planetparam}.  We also used a Gaussian prior on the linear
limb-darkening coefficient to describe the spectroscopic transit, with
central value 0.65 (based on an interpolation of the tables of
\citet{claret:2004}) and a standard deviation of 0.10.  The only
completely free parameters were $\lambda$ and $\gamma$.  We fitted the
58 RVs from 2011~Feb~21, adopting uncertainties equal to the quadrature
sum of the internally-estimated uncertainty and a ``jitter'' term of
4.8~m~s$^{-1}$, the value giving $\chi^2 = N_{\rm dof}$.

\begin{figure*}[ht]
\begin{center}
\leavevmode
\hbox{
\epsfxsize=7.5in
\epsffile{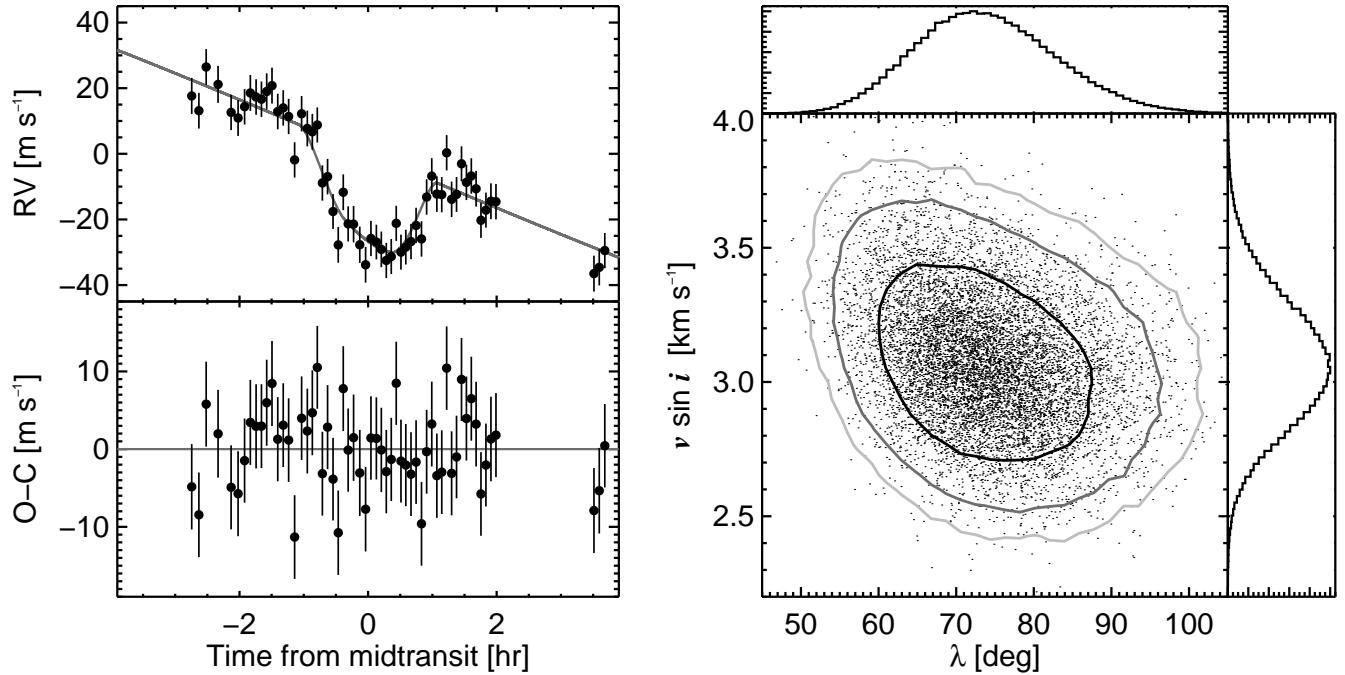}}
\end{center}
\vspace{-0.1in}
\caption{
	{\bf Rossiter-McLaughlin effect for \hatcur{}} {\it
	Left.}---Apparent radial velocity variation on the night of
	2011~Feb~21, spanning a transit.  The top panel shows the observed
	RVs.  The bottom panel shows the residuals between the data and the
	best-fitting model.  {\it Right.}---Joint constraints on $\lambda$
	and $v\sin i_\star$.  The contours represent 68.3\%, 95.4\%, and
	99.73\% confidence limits.  The marginalized posterior probability
	distributions are shown on the sides of the contour plot.
\vspace{0.25in}
\label{fig:hat30}}
\end{figure*}

To derive parameter values and uncertainties we used a Markov Chain
Monte Carlo (MCMC) algorithm employing Gibbs sampling and
Metropolis-Hastings stepping.  \reftabl{planetparam} summarizes the
results for the key parameters.  Figure~\ref{fig:hat30} shows the RV
data and the results for $v\sin i_\star$ and $\lambda$.

The result for $\lambda$ is $\lam \pm \ulam$~degrees. The finding of a
large misalignment is obvious from visual inspection of
Figure~\ref{fig:hat30}, which shows that the anomalous RV was a
blueshift throughout the entire transit, as opposed to the
``red-then-blue'' pattern of a well-aligned system.

The result $v\sin i_\star = 3.07\pm 0.24$~km~s$^{-1}$ based on the RM
analysis is $\sim 2\sigma$ larger than the result $2.2\pm
0.5$~km~s$^{-1}$ that was derived from the observed spectral line
broadening.  This could be a signal of systematic errors in the
spectral analysis, imperfections in our RM calibration scheme, or
possibly even differential rotation (which would affect both
measurements in different ways).  Fortunately the result for $\lambda$
is nearly immune to possible systematic errors in $v\sin i_\star$, as
the errors in those two parameters are nearly
uncorrelated.\footnote{When the prior constraint on $v\sin i_\star$
  was released, and the analysis was repeated, the results were
  $\lambda = 68.9\pm 9.8$~degrees and $v\sin i_\star = 3.39\pm
  0.33$~km~s$^{-1}$.}  This is true of most systems with high transit
impact parameters \citep{gaudi07}.

\section{Discussion}
\label{sec:discussion}

We have reported the discovery of \hatcurb{}, a giant planet in a
close-in orbit around a late F-type dwarf star.  The star is relatively
bright ($V=10.4$), which will facilitate many interesting follow-up
observations, such as studies of the planet's atmosphere through
transmission and occultation spectroscopy.  One such follow-up study,
the detection and analysis of the Rossiter-McLaughlin effect, has
already been conducted and presented in this paper.

The pace of discovery of new exoplanets continues to rise each year,
and the properties of this ever-larger ensemble have revealed many
patterns that hint at the processes shaping the observed architectures
of planetary systems.  One such pattern among the transiting hot
Jupiters was discovered by \citet{winn10} who noted a tendency for
planets around hot stars to have misaligned orbits ($|\lambda| \gtrsim
10^\circ$), while planets orbiting cool stars have orbits that are more
closely aligned with the stellar rotation.  This pattern also manifests
itself in the statistical distribution of projected stellar rotation
rates ($v\sin i_\star$).  \citet{schlaufman10} searched for anomalously
low values of $v\sin i_\star$ among the host stars of transiting
exoplanets, which would be an indication of significant spin-orbit
misalignments along the line of sight (small $\sin i_\star$). 
Schlaufman found that many of the most massive (and hottest) planet
host stars among his sample showed evidence of misalignment.

\hatcurb{} is a newly discovered hot
Jupiter ($P = \hatcurLCP$~days, $\mpl=\hatcurPPmlong\,\mjup$) orbiting
a hot ($T_{\rm eff} = 6300$~K) host star.  Our spectroscopic
observations made during the transit of the planet exhibit the
Rossiter-McLaughlin effect, revealing a sky-projected angle between the
star's spin axis and the planet's orbit normal $\lambda = \lam \pm
\ulam$.  The \hatcur{} system therefore represents another example of a
misaligned hot Jupiter orbiting a hot star, and conforms to the
observed correlation between spin-orbit misalignment and stellar
effective temperature \citep[see also][]{winn11}.  It has also been
suggested that planet mass and orbital eccentricity are factors linked
to spin-orbit misalignment (see, e.g., Johnson et al.~2009, H\'ebrard
et al.~2010), although HAT-P-30 is not particularly massive, nor is the
orbit detectably eccentric.

The reason why hot stars tend to have high obliquities is not known,
but one possible factor is tidal interactions \citep{winn10}.  As
mentioned in the introduction, cool stars have thicker convective
envelopes where tidal dissipation is thought to be stronger.  Close-in
planets might be able to torque the convective layer of cool stars into
alignment, while hot stars lack massive outer convective layers and
their planets remain misaligned.  This scenario suggests that hot
Jupiters generally arrive at their close-in orbits with substantial
orbital tilts.  It also requires a long-lived decoupling between the
convective and radiative layers of Sun-like stars, a situation not
observed in the Sun.  However, the tidal-interaction model makes a
clear prediction: exceptions to the $T_{\rm eff}$--misalignment
correlation should be found in systems containing planets with small
masses or wide orbits, since the strength of tidal interactions is
diminished in either case.  As noted by Winn et al., the only ``strong
exceptions'' among their sample are systems with planets in wide
orbits, including WASP-8b and HD80606b \citep{queloz10,moutou09,
  winn09}. Another test is the case of the HAT-P-11 system, which
contains a Neptune-mass planet ($M_P = 0.1$~\mjup) orbiting a cool
star. Indeed \citet{winn10b} and \citet{hirano11} found that the
planet is in a highly misaligned orbit.

Additional tests are warranted using a larger and more diverse sample
of transiting planets.  Fortunately, this larger sample is forthcoming
since planets around bright ($V < 12$) F- and G-type stars straddling
the proposed division between ``hot'' and ``cold'' ($T_{\rm eff} =
6250$~K) are in the detectability ``sweet spot'' of ground-based
transit surveys such as HATNet.

\acknowledgements 

HATNet operations have been funded by NASA grants NNG04GN74G,
NNX08AF23G and SAO IR\&D grants.  GT acknowledges partial support from
NASA grant NNX09AF59G.  We acknowledge partial support also from the
Kepler Mission under NASA Cooperative Agreement NCC2-1390 (D.W.L., PI). 
G.K.~thanks the Hungarian Scientific Research Foundation (OTKA) for
support through grant K-81373.  This research has made use of Keck
telescope time granted through NASA (N167Hr).  Based in part on data
collected at Subaru Telescope, which is operated by the National
Astronomical Observatory of Japan.



\end{document}